\documentclass[sigconf,natbib=true]{acmart}
\AtBeginDocument{%
  }

\usepackage{listings}
\usepackage{titletoc}
\usepackage{graphicx}
\usepackage[noabbrev,capitalise]{cleveref}

\usepackage{amssymb}
\usepackage{multicol}
\usepackage{multirow}
\usepackage{xspace}
\usepackage{makecell}
\usepackage{bm}
\usepackage{amsmath}
\usepackage{enumitem}
\DeclareMathOperator*{\argmin}{arg\,min}
\usepackage{subcaption}
\usepackage{tablefootnote}

\newcommand{\ignore}[1]{}

\newcommand{\mymodel}{Tlow}

\copyrightyear{2026}
\acmYear{2026}
\setcopyright{cc}
\setcctype{by}
\acmConference[CIKM '26]{Proceedings of the 35th ACM International Conference on Information and Knowledge Management}{November 07--11, 2026}{Rome, Italy}
\acmBooktitle{Proceedings of the 35th ACM International Conference on Information and Knowledge Management (CIKM '26), November 07--11, 2026, Rome, Italy}
\acmDOI{10.1145/3799682.3840087}
\acmISBN{979-8-4007-2539-5/2026/11}
\begin{document}

\title{Tlow: Flow-based Item Tokenizer for Recommendation}
\author{Nian Li}
\authornote{Both authors contribute equally to this work.}
\affiliation{%
  \institution{Tsinghua University}
  \city{Shenzhen}
  \country{China}
}

\author{Chonggang Song}
\authornotemark[1]
\affiliation{%
  \institution{Tencent Inc.}
  \city{Shenzhen}
  \country{China}
}

\author{Jingtao Ding}
\affiliation{%
  \institution{Tsinghua University}
  \city{Beijing}
  \country{China}
}

\author{Lingling Yi}
\affiliation{%
  \institution{Tencent Inc.}
  \city{Shenzhen}
  \country{China}
}

\author{Yong Li}
\affiliation{%
  \institution{Tsinghua University}
  \city{Beijing}
  \country{China}
}

\author{Qingmin Liao}
\affiliation{%
  \institution{Shenzhen International Graduate School, Tsinghua University}
  \city{Shenzhen}
  \country{China}
}

\renewcommand{\shortauthors}{Nian Li et al.}


\begin{abstract}
Item tokenizer encodes semantic embeddings into token IDs to replace the randomly assigned item IDs used in traditional recommendation models, fundamentally addressing the problems of excessive parameters and cold starts. However, the most common tokenizer, RQ-VAE, suffers from low decoding efficiency due to the inherent dependencies among its codebooks. 
Meanwhile, efficient independent tokenizers such as optimized product quantization (OPQ) still struggle with dimensional correlations and distribution complexity of semantic embeddings.
In this work, we propose a f\underline{low}-based item \underline{T}okenizer (Tlow) to transform raw semantic embeddings into a latent space where embeddings conform to a unified standard normal distribution, achieving dual advantages of dimensional independence and distributional simplicity. Independent tokenization performed on these latent embeddings yields semantically clear token IDs. Additionally, we introduce a novel codebook guidance to align the codebook space with the token embedding space, further aiding the learning of more semantically distinct token embeddings. Offline experiments on four public datasets demonstrate that Tlow's tokenization and codebook guidance significantly improve recommendation performance. The improvement on cross-domain and multi-modal recommendations also proves the effectiveness of item tokenization in a simplified embedding space. Online experiments for a multi-modal retrieval task on China's largest social media platform WeChat validate Tlow's powerful distribution transformation capability. The retrieval model based on token IDs improves user CTR by 10.32\% globally and by 11.64\% for new items. Our codes are available at \url{https://github.com/wjjln/Tlow}.
\end{abstract}


\maketitle

\section{Introduction}
Traditional recommender systems often assign a random ID embedding to each item, with learning embeddings as the model's core objective. However, embeddings learned by ID-based models (GNNs~\cite{wang2019neural}, RNNs~\cite{hidasi2015session}, or self-attention~\cite{kang2018self,zhou2018deep}) are completely independent of one another, so the number of parameters grows linearly with the number of items, limiting model capacity in practical online recommender systems. Moreover, the cold-start issue is another significant drawback, \textit{i.e.}, the model struggles to recommend new items with few interaction records.

Inspired by the training paradigm of large language models, some studies now tokenize items based on their semantic embeddings to generate a sequence of discrete token IDs (also called semantic IDs)~\cite{rajput2023recommender,hou2025generating,hou2023learning}, creating a globally shared token vocabulary. Sequential recommendation models then learn token embeddings by decoding the token IDs of the next item. This approach relies on token embeddings to model atomic information in the semantic space of all items, effectively constraining the model's parameter size while naturally resolving the cold-start issue.
A representative work, TIGER~\cite{rajput2023recommender}, uses residual-quantized variational auto-encoder (RQ-VAE)~\cite{zeghidour2021soundstream} as its tokenizer, applying hierarchical encoding to semantic embeddings. This creates strong correlations between codebooks, so decoding the next token ID relies on previously decoded IDs, causing an efficiency bottleneck: the number of inference steps must equal the number of codebooks. Moreover, the codebook collapse and conflicts common in RQ-VAE~\cite{kuai2024breaking} make its industrial application require substantial experience and expertise~\cite{zhou2025onerec}. Pursuing independent tokenization to enable efficient parallel decoding is thus a natural next step, and this decoding-efficiency advantage over RQ-VAE-based sequential decoding has been empirically validated in~\cite{hou2025generating}. Existing methods using product quantization (PQ)~\cite{jegou2010product} directly partition semantic embeddings and encode each part independently~\cite{hou2023learning}, but still face two critical challenges:
\begin{itemize}[leftmargin=*]
    \item \textbf{Dimension correlations}. The correlations between the dimensions of the semantic embeddings violate the dimensional-independence assumption required for independent tokenization. Although optimized product quantization (OPQ)~\cite{ge2013optimized} alleviates this by decomposing the raw space into subspaces via an orthogonal transformation, dimensions within each subspace remain correlated~\cite{hou2025generating}. When dimensions are correlated, embeddings lie on a complex, skewed manifold, but independent quantization imposes a grid-like structure over it, leading to poor quantization~\cite{zhang2022anisotropic}.
    \item \textbf{Complex embedding distribution}. The unknown, complex distribution harms the semantic representational ability of embeddings~\cite{li2020sentence} and causes large quantization errors when tokenized directly. Specifically, these embeddings are often highly anisotropic, occupying a non-uniform cone rather than being evenly spread~\cite{jegou2010product}, making standard quantization codebooks an inefficient fit, especially in cross-domain or multi-modal scenarios where embeddings from diverse sources form distinct, separated clusters.
\end{itemize}

\begin{figure}[t]
    \centering
    \includegraphics[width=0.86\linewidth]{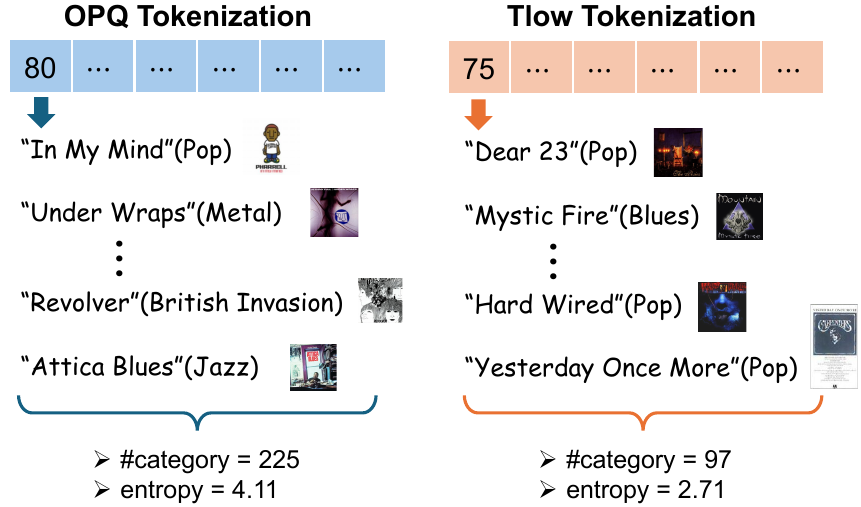}
    \caption{A Comparison of semantic clarity of token IDs generated by OPQ and Tlow tokenization.}
    \label{fig:intro}
\end{figure}

In this work, we propose \textbf{Tlow, a flow-based item tokenizer that transforms raw semantic embeddings into latent embeddings conforming to a standard normal distribution}. Compared to the raw, correlated, and anisotropic embedding space, the transformed latent space better satisfies the dimensional-independence assumption underlying PQ-based independent quantization~\cite{jegou2010product,ge2013optimized}, mitigating the quantization error induced by correlated and skewed distributions. The resulting latent embeddings combine dimensional independence with distributional simplicity, facilitating more accurate tokenization and more semantically distinct token embeddings. Specifically, Tlow processes semantic embeddings with a multi-scale architecture~\cite{kingma2018glow}, built from basic units called ``A Step of Flow'', each comprising three layers: Activation Normalization (ActNorm), Invertible Linear, and Affine Coupling. We apply PQ on the latent embeddings to generate token IDs for each item.
Figure \ref{fig:intro} compares 16-bit (\textit{i.e.}, 16-part) OPQ and Tlow tokenization on the ``CDs and Vinyl'' dataset from Amazon Reviews~\cite{mcauley2015image}, showing albums sharing the same first-position token ID. Although these albums are all related to the ``Pop'' genre, OPQ groups in many albums of different genres~(\textit{e.g.}, ``Metal'', ``British Invasion'', ``Jazz''), whereas Tlow's tokenization yields distinct, clear semantics, with most albums genuinely ``Pop''. Tlow also exhibits significantly lower genre diversity than OPQ, as indicated by fewer categories and lower entropy.
We then use user behavioral data to train the sequential recommendation model and the globally shared token embeddings, introducing a novel codebook guidance that aligns the token embedding space with the codebook space so token embeddings capture distinct semantics and atomic information.

We conduct extensive experiments on four Amazon Reviews datasets from different commodity categories. Tlow significantly improves recommendation performance given the same sequential model, validating the effectiveness of tokenization within a standard normal distribution space; ablation studies further confirm the importance of the latent embeddings' semantic information and the codebook guidance. Performance gains in cross-domain and multi-modal recommendations further support Tlow's effectiveness in simplifying complex semantic embedding distributions, although the improvement is not uniformly largest across every domain and modality metric. On China's largest social media platform WeChat, we conduct online A/B testing for a multi-modal picture\footnote{A typical user post consists of a few pictures and a brief caption.} retrieval task, where replacing item IDs with Tlow's token IDs yields significant gains in conversion metrics, especially for newly published pictures.

\section{\mymodel}
In this work, we formulate the task as sequential recommendations. Specifically, based on each user's historical interaction sequence $\{i_1, i_2, \cdots, i_T\}$, the recommendation model aims to predict the next interaction item $i_{T+1}$, where $T$ is the sequence length. Traditional ID-based models assign a random ID for each item and learn ID embeddings through sequential models like Transformer decoder. However, ID-based models suffer from inherent problems of cold-start issues, excessive parameters, and so on. Item tokenizer leverages the semantic embedding $\mathbf{x}$ to transform each item into a sequence of token IDs, where $\mathbf{x}$ is obtained from pre-trained models with item features~(\textit{e.g.}, title, cover image) as the input. In this way, tokenization-based sequential models have the objective of decoding token IDs of the next item, similar to the training paradigm of large language models. This approach overcomes the aforementioned fundamental problems by learning a shared set of token embeddings.

\subsection{Overview}
As stated in the introduction, dimensional correlations and complex distribution of semantic embeddings lead to challenges of independent tokenization. In this work, we propose \mymodel\ with a multi-scale architecture~\cite{kingma2018glow} to transform semantic embedding $\mathbf{x}\in \mathbb{R}^{d_s}$ into latent embedding $\mathbf{z}\in \mathbb{R}^{d_s}$ conforming to a standard normal distribution, where $d_s$ is the embedding dimension. We can perform more accurate independent tokenization on $\mathbf{z}$, generating token IDs with clearer semantics. 
Like all flow-based models, \mymodel's log-likelihood objective is also to minimize:
\begin{equation}
    \mathcal{L}_f = -\frac{1}{\vert \mathcal{X}\vert}\sum_{\mathbf{x}\in \mathcal{X}}\log p_{\boldsymbol{\theta}}(\mathbf{x}),
\end{equation}
where $\mathcal{X}$ denotes the set of semantic embeddings of all the items and $\boldsymbol{\theta}$ are learnable model parameters. The probability density function of \mymodel\ can be further written as:
\begin{equation}
    \log p_{\boldsymbol{\theta}}(\mathbf{x}) = \log p_{\boldsymbol{\theta}}(\mathbf{z}) + \log\vert\det(\text{d}\mathbf{z}/\text{d}\mathbf{x})\vert,
\end{equation}
where $\text{d}\mathbf{z}/\text{d}\mathbf{x}$ is the Jacobian matrix of the transformation from $\mathbf{x}$ to $\mathbf{z}$, and $\det(\cdot)$ denotes the matrix determinant.

\begin{figure*}[t]
    \centering
    \includegraphics[width=0.9\linewidth]{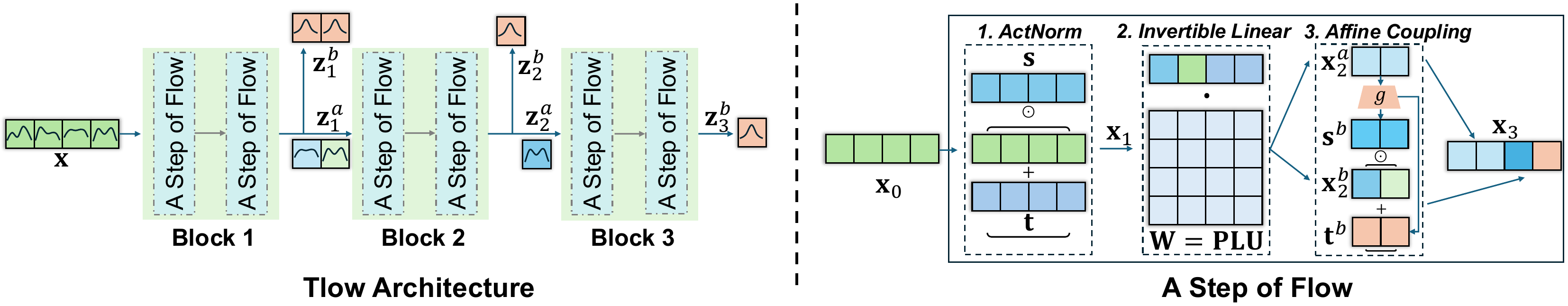}
    \caption{The illustration of \mymodel\ Architecture with $N=3$ blocks and $M=2$ flows, where the dimension of semantic embedding is $d_s = 4$ (left). An example of ``A Step of Flow'' processing embedding with the dimension $d=4$ (right).}
    \label{fig:tlow}
\end{figure*}

\mymodel\ contains multiple blocks, each consisting of a multi-step flow where each step includes three transformation layers, \textit{i.e.}, ActNorm, Invertible Linear, and Aﬃne Coupling layer. Figure \ref{fig:tlow} illustrates the model architecture of \mymodel. 

\subsection{A Step of Flow}
A step of flow consists of the following three transformation layers:
\begin{itemize}[leftmargin=*]
\item{\textbf{ActNorm Layer}.}
For the input $\mathbf{x}_0\in \mathbb{R}^{d}$, this layer performs activation normalization for stable training as follows,
\begin{equation}
    \mathbf{x}_1 = \mathbf{s}\odot (\mathbf{x}_0 + \mathbf{t}),
\end{equation}
where $\odot$ denotes the element-wise product and $\mathbf{s}, \mathbf{t}\in \mathbb{R}^{d}$ are learnable scale and bias parameters, respectively. The log-determinant of this layer is:
\begin{equation}
    \log\vert\det(\text{d}\mathbf{x}_1/\text{d}\mathbf{x}_0)\vert = \sum_{j=1}^d\log\vert \mathbf{s}_j \vert.
\end{equation}

\item{\textbf{Invertible Linear Layer}.}
For the input $\mathbf{x}_1$, this layer learns an invertible linear transformation to allow the dimensions of the embedding to influence one another:
\begin{equation}
    \mathbf{x}_2 = \mathbf{W}\mathbf{x}_1 = (\mathbf{P}\mathbf{L}\mathbf{U})\mathbf{x}_1.
\end{equation}
The linear matrix $\mathbf{W}\in \mathbb{R}^{d\times d}$ is decomposed into the product of a permutation matrix $\mathbf{P}$, a lower triangular matrix $\mathbf{L}$, and an upper triangular matrix $\mathbf{U}$ to facilitate the computation of the determinant. We keep the diagonal elements of $\mathbf{L}$ as ones, while making the diagonal of $\mathbf{U}$ a set of learnable parameters $\mathbf{w}\in \mathbb{R}^d$. In this way, the log-determinant of this layer is:
\begin{equation}
\begin{aligned}
    \log\vert\det(\text{d}\mathbf{x}_2/\text{d}\mathbf{x}_1)\vert &= \log\vert\det(\mathbf{W})\vert \\
    &= \log\vert\det(\mathbf{P})\vert + \log\vert\det(\mathbf{L})\vert + \log\vert\det(\mathbf{U})\vert \\
    &= \sum_{j=1}^d\log\vert \mathbf{w}_j \vert.
\end{aligned}
\end{equation}

\item{\textbf{Aﬃne Coupling Layer}.}
For the input $\mathbf{x}_2$, this layer first divides it into two halves $\mathbf{x}_2^a$ and $\mathbf{x}_2^b$, and performs the transformation as follows:
\begin{equation}
\begin{aligned}
    &\mathbf{x}_3 = [\mathbf{x}_2^a, \mathbf{s}^b\odot (\mathbf{x}_2^b + \mathbf{t}^b)], \\
    &[\mathbf{s}^b, \mathbf{t}^b] = g(\mathbf{x}_2^a),
\end{aligned}
\end{equation}
where $g: \mathbb{R}^{\frac{d}{2}} \rightarrow \mathbb{R}^d$ is a MLP to generate scale and bias parameters based on $\mathbf{x}_2^a$ to transform $\mathbf{x}_2^b$. The log-determinant of this layer is:
\begin{equation}
    \log\vert\det(\text{d}\mathbf{x}_3/\text{d}\mathbf{x}_2)\vert = \sum_{j=1}^{d/2}\log\vert \mathbf{s}_j^b \vert.
\end{equation}
\end{itemize}
Combining these three layers together, the log-determinant of a step of flow is:
\begin{equation}
    \Delta = \log\vert\det(\text{d}\mathbf{x}_1/\text{d}\mathbf{x}_0)\vert + \log\vert\det(\text{d}\mathbf{x}_2/\text{d}\mathbf{x}_1)\vert + \log\vert\det(\text{d}\mathbf{x}_3/\text{d}\mathbf{x}_2)\vert.
\end{equation}

\subsection{Flow-based Tokenizer}
\mymodel\ contains $N$ blocks and each block consists of $M$ steps of flow. For the $n$-th block, the output $\mathbf{z}_n\in \mathbb{R}^{\frac{d_s}{2^{n-1}}}$ is divided into two halves $\mathbf{z}_n^a\in \mathbb{R}^{\frac{d_s}{2^{n}}}$ and $\mathbf{z}_n^b\in \mathbb{R}^{\frac{d_s}{2^{n}}}$, which serve as the input of the next block and partial output~(\textit{i.e.}, partial $\mathbf{z}$) of \mymodel, respectively. We estimate the log-likelihood of $\mathbf{z}_n^b$ as follows:
\begin{equation}
\begin{aligned}
    &\log p(\mathbf{z}_n^b) = \log \mathcal{N}(\mathbf{z}_n^b; \boldsymbol{\mu}_n, \boldsymbol{\sigma}_n^2), \\
    &[\boldsymbol{\mu}_n, \boldsymbol{\sigma}_n^2] = h(\mathbf{z}_n^a),
\end{aligned}
\end{equation}
where $h: \mathbb{R}^{\frac{d_s}{2^n}} \rightarrow \mathbb{R}^{\frac{d_s}{2^{n-1}}}$ is an MLP module to predict the mean $\boldsymbol{\mu}_n$ and variance $\boldsymbol{\sigma}_n^2$ of $\mathbf{z}_n^b$ based on $\mathbf{z}_n^a$. Combining the outputs of all the blocks, we obtain final output of \mymodel\ $\mathbf{z} = [\mathbf{z}_1^b, \mathbf{z}_2^b, \cdots, \mathbf{z}_N^b] \in \mathbb{R}^{d_s}$. Note that $\mathbf{z}_N^b = \mathbf{z}_N$ since we do not need to divide the output of the last block. Moreover, the input of the first block is $\mathbf{x}$. We then formulate:
\begin{equation}
    \log p_{\boldsymbol{\theta}}(\mathbf{z}) = \sum_{n=1}^N \log p(\mathbf{z}_n^b).
\end{equation}
In addition, the log-determinant of \mymodel\ is calculated as follows:
\begin{equation}
    \log\vert\det(\text{d}\mathbf{z}/\text{d}\mathbf{x})\vert = \sum_{n=1}^N \sum_{m=1}^M \Delta_{n,m},
\end{equation}
where $\Delta_{n,m}$ denotes the log-determinant of the $m$-th step of flow in the $n$-th block. After training \mymodel\ with the log-likelihood objective $\mathcal{L}_f$, the semantic embedding $\mathbf{x}$ is transformed into a latent embedding $\mathbf{z}$ conforming to a standard normal distribution, where the dimensions are independent of each other.

For item tokenization, we perform the PQ on $\mathbf{z}$ to obtain $C$ token IDs. Specifically, $\mathbf{z}$ is divided into $C$ parts $[\mathbf{z}_1, \mathbf{z}_2, \cdots, \mathbf{z}_C]$, each of which is independently encoded using K-means to obtain a corresponding codebook $\mathcal{C}_k = \{\mathbf{c}_k^1, \mathbf{c}_k^2, \cdots, \mathbf{c}_k^S\}$. Here, $\mathbf{c}_k^j\in \mathbb{R}^{\frac{d_s}{C}}$ is the $j$-th quantized codeword and $S$ denotes the codebook size. For the $k$-th tokenization, the token ID is defined as:
\begin{equation}
    c_k = \argmin_{j\in \{1, 2, \cdots, S\}}\Vert \mathbf{z}_k - \mathbf{c}_k^j\Vert^2.
\end{equation}
In this way, \mymodel\ tokenizes each item into discrete IDs $[c_1, c_2, \cdots, c_C]$.

\subsection{Token IDs for Recommendation}
Given token IDs of all the items, a sequential model (\textit{e.g.}, Transformer decoder) aims to generate the IDs of the next item, with a set of token embeddings learned. Specifically, there is a one-to-one correspondence between the token embeddings $\mathbf{E}\in \mathbb{R}^{C\times S\times d_m}$ and codebooks  $\mathbf{C}\in \mathbb{R}^{C\times S\times \frac{d_s}{C}}$, where $d_m$ denotes the dimension of token embeddings. Traditional ID embedding can be replaced with the aggregated token embeddings $\mathbf{e}\in \mathbb{R}^{d_m}$ for each item:
\begin{equation}
    \mathbf{e} = \frac{1}{C}\sum_{k=1}^C \mathbf{E}_{k, c_k}.
\end{equation}
The sequential model further encodes the user's historical embedding sequence $\left[\mathbf{e}_{i_1}, \mathbf{e}_{i_2}, \cdots\right]$ and outputs a final hidden state $\mathbf{h}\in \mathbb{R}^{d_m}$. Following \cite{hou2025generating}, we predict the log-likelihood of interaction with the target item $i_t$ with token IDs $\left[c_1^t, c_2^t, \cdots, c_C^t\right]$ as follows:
\begin{equation}
    \label{eq:pmodel}
    \log p(i_t\vert \mathbf{h}) = \sum_{k=1}^C\log p\left(c_k^t\vert \mathbf{h}\right) = \sum_{k=1}^C\log \frac{\exp{\left(\mathbf{E}_{k, c_k^t}^\top g_k(\mathbf{h})/\tau\right)}}{\sum_{j=1}^S\exp{\left(\mathbf{E}_{k, j}^\top g_k(\mathbf{h})/\tau\right)}},
\end{equation}
where $\tau$ is the temperature hyperparameter. The $k$-th project head $g_k: \mathbb{R}^{d_m} \rightarrow \mathbb{R}^{d_m}$ decodes the $k$-th token ID of the next item. This decoding approach owes to the independence between different parts of the latent embedding $\mathbf{z}$. Furthermore, the training loss for recommendation can be formulated as:
\begin{equation}
    \mathcal{L}_{rec} = -\sum_{(\mathbf{h}, i_t)} \log p(i_t\vert \mathbf{h}).
\end{equation}

\subsection{Learning Token Embeddings with Codebook Guidance}
Since \mymodel\ tokenizes items based on latent embeddings, the quantized codebooks imply clearer semantics contained in the whole set of items. In other words, each codeword in a codebook encodes atomic information derived from the decomposition of items' semantics, as the illustration in Figure \ref{fig:intro}. Therefore, learning token embeddings with the codebook guidance will help enhance recommendations. Unlike independent alignment for each item or behavior sequence used in existing works~\cite{wang2024learnable,liu2024generative}, we propose to directly align two latent spaces spanned by token embeddings and codebooks. The token space $\mathbf{\Phi}\in \mathbb{R}^{C\times S\times S}$ and codebook space $\mathbf{\Psi}\in \mathbb{R}^{C\times S\times S}$ are defined based on cosine similarities between token embeddings and codewords as follows:
\begin{equation}
    \mathbf{\Phi}_{k,i,j} = \frac{\mathbf{E}_{k, i}^\top \mathbf{E}_{k, j}}{\Vert\mathbf{E}_{k, i}\Vert\cdot \Vert\mathbf{E}_{k, j}\Vert},
    \mathbf{\Psi}_{k,i,j} = \frac{\mathbf{C}_{k, i}^\top \mathbf{C}_{k, j}}{\Vert\mathbf{C}_{k, i}\Vert\cdot \Vert\mathbf{C}_{k, j}\Vert}.
\end{equation}
Furthermore, we implement the codebook guidance as a simple but effective MSE loss:
\begin{equation}
    \mathcal{L}_{sim} = \text{MSE}\left(\vert\mathbf{\Phi} - \mathbf{\Psi}\vert\right).
\end{equation}
Finally, the sequential model and token embeddings are learned by training the combined loss $\mathcal{L} = \mathcal{L}_{rec} + \lambda\mathcal{L}_{sim}$, where $\lambda$ is the degree of codebook guidance.

\section{Experiments}
We conduct experiments under three scenarios to evaluate the effectiveness of our proposed \mymodel, including general, cross-domain, and multi-modal sequential recommendation.
\subsection{Experimental Setup}
\subsubsection{Datasets}
We use four commodity categories in Amazon Reviews dataset~\cite{mcauley2015image} for experiments, including ``Sports and Outdoors (Sports),
``Beauty'', ``Toys and Games (Toys)'', and ``CDs and Vinyl
(CDs)''. Following existing works~\cite{hou2025generating,hou2023learning,rajput2023recommender}, users' reviews are regarded as item interactions and arranged chronologically as historical item sequences. We also use a processed dataset ``Cloth-Sports'' from a state-of-the-art~(SOTA) work~\cite{liu2025bridge} for cross-domain sequential recommendation. This dataset is collected from ``Clothing Shoes and Jewelry'' and ``Sports and Outdoors'' categories in Amazon dataset. Most of the users in this dataset overlap across two categories. The statistics of used datasets are shown in Table \ref{tab:dataset}, where Avg. Length means the length of historical sequence averaged over all the users.

The last item in a historical sequence is used for testing and the second-to-last item for validation, which is the widely adopted leave-one-out evaluation protocol in existing works~\cite{kang2018self,hou2025generating,zhao2022revisiting}.

\begin{table}[!t] %
  \centering
\caption{Statistics of used datasets.}
\label{tab:dataset}
\begin{tabular}{ccccc}
  \toprule
  \textbf{Datasets} & \textbf{\#Users} & \textbf{\#Items} & \textbf{\#Interactions} & \textbf{Avg. Length}\\
  \midrule
  \textbf{Sports}  & 35,598             & 18,357           & 260,739            & 8.32 \\
  \textbf{Beauty}  & 22,363            & 12,101           & 176,139            & 8.87 \\
  \textbf{Toys}    & 19,412            & 11,924           & 148,185            & 8.63 \\
  \textbf{CDs}     & 75,258            & 64,443           & 1,022,334          & 14.58 \\
  \midrule
  \textbf{Cloth} & 9,933 & 3,278 & 97,741 & \multirow{2}{*}{10.71} \\
  \textbf{Sports} & 4,263 & 1,021 & 11,879 & \\
  \bottomrule
\end{tabular}
\end{table}

\subsubsection{Baselines}
We compare \mymodel\ with two types of baselines following existing works~\cite{hou2025generating,liu2024generative,rajput2023recommender}. \textbf{\textit{Traditional ID-based models}} include Caser~\cite{tang2018personalized}, GRU4Rec~\cite{hidasi2015session}, HGN~\cite{ma2019hierarchical}, BERT4Rec~\cite{sun2019bert4rec}, SASRec~\cite{kang2018self}, FDSA~\cite{zhang2019feature}, and S$^3$-Rec~\cite{zhou2020s3}, which adopt different neural architectures~(\textit{e.g.}, CNNs, RNNs, and Transformers) to model user behavior sequences with randomly assigned item IDs. \textbf{\textit{Tokenization-based models}} include VQRec~\cite{hou2023learning}, TIGER~\cite{rajput2023recommender}, ETEGRec~\cite{liu2024generative}, RecJPQ~\cite{petrov2024recjpq}, HSTU~\cite{zhai2024actions}, and RPG~\cite{hou2025generating}, which leverage various quantization techniques~(\textit{e.g.}, PQ, RQ-VAE, and OPQ) to tokenize item semantic embeddings for generative recommendation.

\subsubsection{Evaluation Metrics}
We use widely adopted ranking metrics Recall@$k$~(R@$k$) and NDCG@$k$~(N@$k$) for performance evaluation, where $k$ is set as $5$ and $10$ following~\cite{hou2025generating,rajput2023recommender}. All the items are included for full ranking to avoid sampling bias when calculating evaluation metrics~\cite{krichene2020sampled}.

\subsubsection{Implementation Details}
We implement our \mymodel\ with Pytorch. All the experimental setups for training and evaluation follow RPG~\cite{hou2025generating}, thus we use baseline results from the original paper and rerun RPG. For a fair comparison, all the models use the extracted item semantic embeddings by the text encoder \texttt{sentence-t5-base}~\cite{ni2021sentence}, leading to $d_s = 768$. We set the number of blocks and the steps of flow in \mymodel\ as $N = 4$ and $M = 4$. In fact, the recommendation performance is not sensitive to the number of blocks and flows, because Tlow can easily transform the semantic embeddings into latent embeddings conforming to a standard normal distribution. The number of codebooks $C\in \{16, 32, 64, 96, 128\}$ and the codebook size $S$ is fixed as $256$ following RPG. The temperature $\tau\in \{0.03, 0.05, 0.07\}$ when decoding token IDs. For the hyperparameters of the sequential model, we keep them consistent with those of RPG to ensure a similar number of model parameters: a 2-layer GPT-2 with the token embedding dimension of $d_m = 448$. Since \mymodel\ decodes all $C$ token IDs of the next item independently within a single forward pass (Eq. \ref{eq:pmodel}), in the same manner as RPG, it retains the same decoding parallelism as RPG and thus the same decoding-efficiency advantage over RQ-VAE-based sequential decoding~(\textit{e.g.}, TIGER) that has already been empirically validated in~\cite{hou2025generating}; we therefore do not repeat separate latency/throughput benchmarking in this work.

\subsection{Overall Performance}
The overall performance comparison is shown in Table \ref{tab:overall}. It is obvious that tokenization-based models using item semantics significantly outperform purely ID-based models. These results prove the effectiveness of item tokenization for more accurate modeling of user behaviors. Moreover, \mymodel\ performs best on all the metrics across four datasets, demonstrating better tokenization in the transformed latent space through a flow-based model. It is worth noting that when using the same semantic embeddings generated by \texttt{sentence-t5-base}, RPG with its independent tokenizer does not outperform TIGER consistently. This also demonstrates the crucial role of Tlow's tokenization on latent embeddings conforming to a standard normal distribution.

\begin{table*}[!t]
\renewcommand{\arraystretch}{1.09}
  \small
    \centering
    \caption{Performance comparison on general recommendation. \underline{Underline} denotes the best baseline and bold denotes better performance than the best baseline at a significance level of $p<0.01$ under paired t-test.
  }
    \label{tab:overall}
  \setlength{\tabcolsep}{1.0mm}{
    \begin{tabular}{cccccccccccccccccc}
      \toprule
      & \multicolumn{1}{c}{\multirow{2.5}{*}{\textbf{Model}}} & \multicolumn{4}{c}{\textbf{Sports and Outdoors}} & \multicolumn{4}{c}{\textbf{Beauty}} & \multicolumn{4}{c}{\textbf{Toys and Games}} & \multicolumn{4}{c}{\textbf{CDs and Vinyl}} \\
      \cmidrule(l){3-6} \cmidrule(l){7-10}\cmidrule(l){11-14}\cmidrule(l){15-18}
      & & R@5 & N@5 & R@10 & N@10 & R@5 & N@5 & R@10 & N@10 & R@5 & N@5 & R@10 & N@10 & R@5 & N@5 & R@10 & N@10 \\
      \midrule
      \multirow{7}{0.5em}{\rotatebox[origin=c]{90}{\textit{ID}}}
      & Caser & 0.0116 & 0.0072 & 0.0194 & 0.0097 & 0.0205 & 0.0131 & 0.0347 & 0.0176 & 0.0166 & 0.0107 & 0.0270 & 0.0141 & 0.0116 & 0.0073 & 0.0205 & 0.0101 \\
& GRU4Rec & 0.0129 & 0.0086 & 0.0204 & 0.0110 & 0.0164 & 0.0099 & 0.0283 & 0.0137 & 0.0097 & 0.0059 & 0.0176 & 0.0084 & 0.0195 & 0.0120 & 0.0353 & 0.0171 \\
& HGN & 0.0189 & 0.0120 & 0.0313 & 0.0159 & 0.0325 & 0.0206 & 0.0512 & 0.0266 & 0.0321 & 0.0221 & 0.0497 & 0.0277 & 0.0259 & 0.0153 & 0.0467 & 0.0220 \\
& BERT4Rec & 0.0115 & 0.0075 & 0.0191 & 0.0099 & 0.0203 & 0.0124 & 0.0347 & 0.0170 & 0.0116 & 0.0071 & 0.0203 & 0.0099 & 0.0326 & 0.0201 & 0.0547 & 0.0271 \\
& SASRec & 0.0233 & 0.0154 & 0.0350 & 0.0192 & 0.0387 & 0.0249 & 0.0605 & 0.0318 & 0.0463 & 0.0306 & 0.0675 & 0.0374 & 0.0351 & 0.0177 & 0.0619 & 0.0263 \\
& FDSA & 0.0182 & 0.0122 & 0.0288 & 0.0156 & 0.0267 & 0.0163 & 0.0407 & 0.0208 & 0.0228 & 0.0140 & 0.0381 & 0.0189 & 0.0226 & 0.0137 & 0.0378 & 0.0186 \\
& S$^3$-Rec & 0.0251 & 0.0161 & 0.0385 & 0.0204 & 0.0387 & 0.0244 & 0.0647 & 0.0327 & 0.0443 & 0.0294 & 0.0700 & 0.0376 & 0.0213 & 0.0130 & 0.0375 & 0.0182 \\
      \midrule
      \multirow{6}{0.5em}{\rotatebox[origin=c]{90}{\textit{Tokenization}}}
     &  RecJPQ & 0.0141 & 0.0076 & 0.0220 & 0.0102 & 0.0311 & 0.0167 & 0.0482 & 0.0222 & 0.0331 & 0.0182 & 0.0484 & 0.0231 & 0.0075 & 0.0046 & 0.0138 & 0.0066 \\
& VQ-Rec & 0.0208 & 0.0144 & 0.0300 & 0.0173 & 0.0457 & 0.0317 & 0.0664 & 0.0383 & 0.0497 & 0.0346 & 0.0737 & 0.0423 & 0.0352 & 0.0238 & 0.0520 & 0.0292 \\
& TIGER & 0.0264 & 0.0181 & 0.0400 & 0.0225 & 0.0454 & 0.0321 & 0.0648 & 0.0384 & \underline{0.0521} & \underline{0.0371} & 0.0712 & 0.0432 & \underline{0.0492} & \underline{0.0329} & \underline{0.0748} & \underline{0.0411} \\
& ETEGRec & 0.0175 & 0.0114 & 0.0281 & 0.0149 & 0.0404 & 0.0277 & 0.0587 & 0.0337 & 0.0209 & 0.0136 & 0.0339 & 0.0178 & 0.0309 & 0.0204 & 0.0461 & 0.0253 \\
& HSTU & 0.0258 & 0.0165 & 0.0414 & 0.0215 & 0.0469 & 0.0314 & 0.0704 & 0.0389 & 0.0433 & 0.0281 & 0.0669 & 0.0357 & 0.0417 & 0.0275 & 0.0638 & 0.0346 \\
      
      & RPG & \underline{0.0296} & \underline{0.0203} & \underline{0.0428} & \underline{0.0246} & \underline{0.0533} & \underline{0.0366} & \underline{0.0753} & \underline{0.0437} & 0.0509 & 0.0357 & \underline{0.0765} & \underline{0.0440} & 0.0486 & 0.0328 & 0.0693 & 0.0395 \\
& \textbf{\mymodel} & \textbf{0.0307} & \textbf{0.0207} & \textbf{0.0477} & \textbf{0.0261} & \textbf{0.0545} & \textbf{0.0377} & \textbf{0.0786} & \textbf{0.0454} & \textbf{0.0590} & \textbf{0.0395} & \textbf{0.0864} & \textbf{0.0482} & \textbf{0.0541} & \textbf{0.0362} & \textbf{0.0801} & \textbf{0.0446} \\
\hline
& \textit{Impr.} (\%) & 3.72&1.97&11.45&6.10&2.25&3.01&4.38&3.89&11.52 & 9.70 & 9.80 & 11.36 & 9.96 & 10.03 & 7.09 & 8.52 \\
      \bottomrule
    \end{tabular}}
  \end{table*}

\subsection{Ablation Study}
We conduct further experiments to validate the effectiveness of two critical modules in \mymodel: the transformed latent embeddings $\mathbf{z}$ and the codebook guidance. 
\begin{itemize}[leftmargin=*]
    \item \textbf{Random} $\mathbf{z}$. Latent embeddings $\mathbf{z}$ are sampled directly from a standard normal distribution instead of being transformed from semantic embeddings by \mymodel. 
    \item \textbf{w/o} $\mathcal{L}_{sim}$. The codebook guidance is removed by setting its corresponding degree $\lambda = 0$.
\end{itemize}
The results in Table \ref{tab:ablation} demonstrate the effectiveness of codebook guidance for enhanced recommendation through learning a semantically clearer latent space of token embeddings. Furthermore, the tokenization based on randomly sampled $\mathbf{z}$ leads to a significant performance drop. This confirms that \mymodel\ retains rich semantic information necessary for behaviors modeling when transforming the embedding $\mathbf{x}$.

\begin{table}[!t]
  \small
    \centering
    \caption{Ablation study of \mymodel.}
    \label{tab:ablation}
  \setlength{\tabcolsep}{0.8mm}{
    \begin{tabular}{ccccccccc}
      \toprule
\multicolumn{1}{c}{\multirow{2.5}{*}{\textbf{Model}}} & \multicolumn{2}{c}{\textbf{Sports}} & \multicolumn{2}{c}{\textbf{Beauty}} & \multicolumn{2}{c}{\textbf{Toys}} & \multicolumn{2}{c}{\textbf{CDs}} \\
      \cmidrule(l){2-3} \cmidrule(l){4-5}\cmidrule(l){6-7} \cmidrule(l){8-9}
    & R@10 & N@10 & R@10 & N@10 & R@10 & N@10 & R@10 & N@10  \\
      \midrule
      \textbf{\mymodel} & \textbf{0.0477} & \textbf{0.0261} & \textbf{0.0786} & \textbf{0.0454} & \textbf{0.0864} & \textbf{0.0482} & \textbf{0.0801} & \textbf{0.0446}\\
      w/o $\mathcal{L}_{sim}$ &0.0449	&0.0248  &0.0764	&0.0436 &0.0826	&0.0471 &0.0756	&0.0425\\
      Random $\mathbf{z}$ &0.0202	&0.0103&0.0623	&0.0363&0.0571	&0.0341&0.0147	&0.0074 \\
\bottomrule
    \end{tabular}}
  \end{table}

\subsection{Cross-domain and Multi-modal Recommendation}
In this section, we further validate the necessity and effectiveness of transforming semantic embeddings from different distributional spaces into a unified space for tokenization. 

\subsubsection{Cross-domain Recommendation}
We choose the current SOTA cross-domain model LLM4CDSR~\cite{liu2025bridge} as our baseline, which leverages LLMs to bridge the domain gap with item semantic embeddings and hierarchical user profiling. Besides, RPG is also included to demonstrate the effectiveness of \mymodel\ when tokenizing items from different domains. For the implementation of RPG and \mymodel, we directly use the mixed item sequences merged from two domains for model training. 

The results in Table \ref{tab:cross-domain} demonstrates the significant improvement of \mymodel\ over both LLM4CDSR and RPG. It is effective for two reasons: first, tokenizing items to create a shared vocabulary of token IDs and training their embeddings naturally bridges the cross-domain gap. Second, Tlow's ability to transform embedding distributions is highly effective for improving recommendations across different domains.

\begin{table}[!t]
    \centering
    \caption{Performance comparison on cross-domain recommendation.}
    \label{tab:cross-domain}
  \setlength{\tabcolsep}{1.2mm}{
    \begin{tabular}{ccccccc}
      \toprule
\multicolumn{1}{c}{\multirow{2.5}{*}{\textbf{Model}}} & \multicolumn{2}{c}{\textbf{Overall}} & \multicolumn{2}{c}{\textbf{Cloth}} & \multicolumn{2}{c}{\textbf{Sports}} \\
      \cmidrule(l){2-3} \cmidrule(l){4-5}\cmidrule(l){6-7}
       & R@10 & N@10 & R@10 & N@10 & R@10 & N@10  \\
      \midrule
      LLM4CDSR & 0.4620 & 0.2803 & 0.4220 & 0.2637 & \textbf{0.5507} & 0.3172  \\
      RPG & 0.4766&	0.3457&	0.4487&	0.3353&	0.5385&	0.3686  \\
      \mymodel\ &\textbf{0.5558}&	\textbf{0.4395}&	\textbf{0.5669}&	\textbf{0.4568}&	0.5312&	\textbf{0.4014} \\
\bottomrule
    \end{tabular}}
  \end{table}
  
\subsubsection{Multi-modal Recommendation}
We chose the current SOTA multi-modal model HM4SR~\cite{zhang2025hierarchical} as our baseline, which leverages interactive and temporal mixture of experts to capture user dynamic interests. Since HM4SR utilizes a large amount of additional side-information~(including interaction timestamp, item category, semantic embeddings as additional input), we replace the ID embeddings in HM4SR with the token embeddings $\mathbf{e}$ generated by Tlow and RPG to directly compare the effectiveness of tokenization.

We conduct experiments on the ``Sports'' dataset and generate additional image embeddings using the advanced CLIP model \texttt{OpenCLIP ViT-H/14}~\cite{radford2021learning}, where the image is downloaded from product URLs of each item. The tokenization on text and image embeddings are performed independently, and both token IDs are merged as final token IDs. The results in Table \ref{tab:multi-modal} show that RPG's tokenization brings limited and inconsistent gains over HM4SR in this setting with combined text-image distributions (\textit{e.g.}, it ties HM4SR on R@5 and underperforms on N@5 and N@10). In contrast, \mymodel\ consistently improves over both baselines on all four metrics after performing the distribution transformation, although the absolute margin over HM4SR (\textit{e.g.}, R@10 from 0.0469 to 0.0521) is moderate in this single-domain multi-modal setting; we expect larger gains with richer or more heterogeneous modalities.

\begin{table}[!t]
    \centering
    \caption{Performance comparison on multi-modal recommendation.}
    \label{tab:multi-modal}
  \setlength{\tabcolsep}{1.2mm}{
    \begin{tabular}{ccccc}
      \toprule
       Model & R@5 & N@5 & R@10 & N@10  \\
      \midrule
      HM4SR & 0.0326&	0.0231&	0.0469&	0.0277 \\
      RPG & 0.0326&	0.0216&	0.0501&	0.0272  \\
      \textbf{\mymodel}\ &\textbf{0.0343} & \textbf{0.0233}&	\textbf{0.0521} &\textbf{0.0290} \\
\bottomrule
    \end{tabular}}
  \end{table}

\subsection{Cold-Start Recommendation}
\label{subsec:coldstart}
Focusing on item semantics, tokenization-based models inherently have an advantage in cold-start recommendation. Therefore, we further investigate whether \mymodel's tokenization can enhance recommendation performance in cold-start scenarios. We conduct our validation from both user and item perspectives. Specifically, users are grouped by their number of historical interactions (interaction depth), while items in the test set are grouped by their frequency in the training set (popularity). We then evaluate the recommendation performance for these different groups on the largest ``CDs'' dataset.

\begin{figure}[!t]
    \centering
    \includegraphics[width=0.49\linewidth]{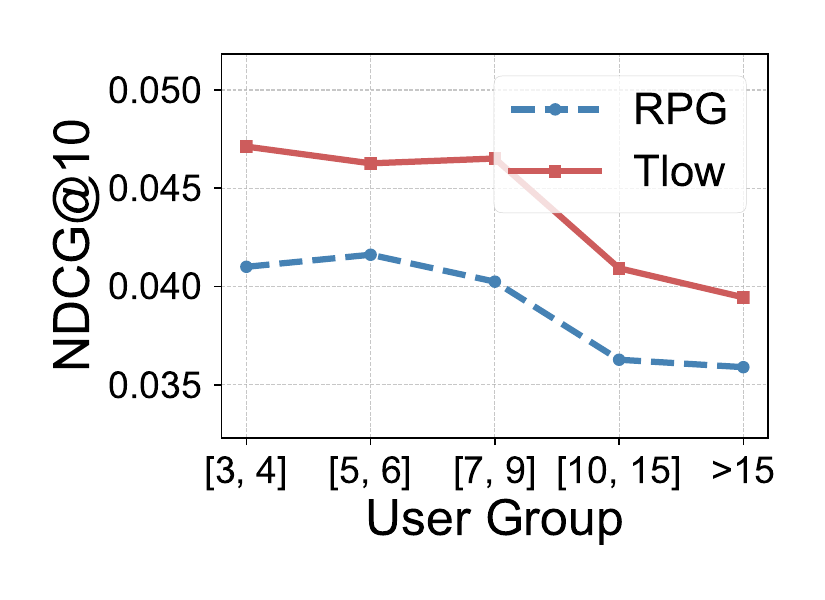}
    \includegraphics[width=0.49\linewidth]{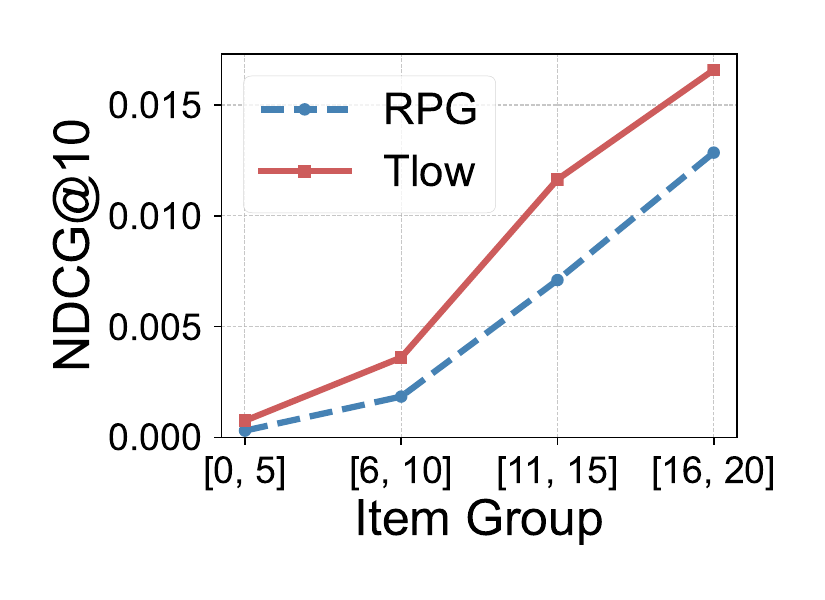}
    \caption{Performance comparison under different user and item groups.}
    \label{fig:cold}
\end{figure}

Figure \ref{fig:cold} shows the performance comparison between \mymodel\ and RPG, where \mymodel\ performs significantly better across all the groups. It indicates that due to \mymodel's ability to transform any embeddings (including long-tail ones) into a standard normal distribution, its tokenization on items across all popularity levels is more comprehensive. This further leads to more accurate behavioral modeling for users with varying interaction depths.

\subsection{Online Experiments}
\subsubsection{Online Setup}
\begin{figure}[!t]
    \centering
    \includegraphics[width=0.92\linewidth]{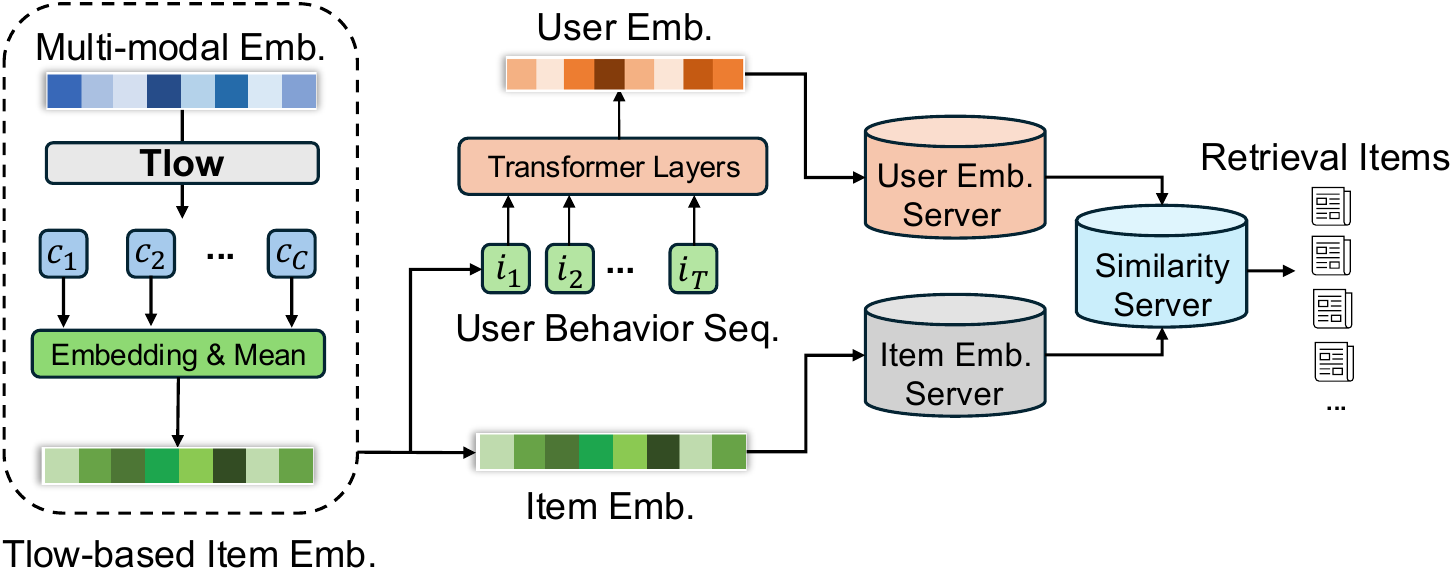}
    \caption{Online serving of Tlow's tokenization.}
    \label{fig:ind}
\end{figure}

We verify Tlow's online performance on China's largest social media platform WeChat, in the picture recommendation scenario involving image-text multi-modal features, comparing sequential models that use Tlow's tokenized embeddings $\mathbf{e}$ against those using randomly assigned item-ID embeddings. We construct users' behavior sequences of length 500 with their recently clicked items, and feed the sequence of item embeddings into a 12-layer Transformer decoder to predict the user's interaction tendency toward the next item. Both sequential models are integrated as retrieval pathways in addition to the primary DSSM-based real-time retrieval method, and the online serving scheme is shown in Figure \ref{fig:ind}. During the online serving phase, user embeddings and item embeddings are stored in dedicated embedding servers. Upon user access, the corresponding user embedding is retrieved via a real-time lookup operation, and candidate pictures with the highest similarity scores are then dynamically identified by the Similarity Server~\cite{faiss2024} to construct the retrieval results. We conduct both single-domain and cross-domain online experiments:
\begin{itemize}[leftmargin=*]
    \item \textbf{Single-domain}: Only users' clicked pictures are used to construct behavior sequences for model training.
    \item \textbf{Cross-domain}: Both users' clicked pictures and articles are mixed to construct behavior sequences for model training. This is to verify the effectiveness of Tlow's tokenization in cross-domain scenarios.
\end{itemize}

\subsubsection{Training and Inference}
In the stage of training Tlow, we find that the model converges after training on embeddings of approximately 1 million items (pictures and articles). Then Tlow can generate token IDs for other unseen items by performing semantic tokenization. As a result, the training time of Tlow is negligible, and the inference efficiency is such that tokenization for tens of millions of newly published items per day can be completed in just a few minutes on a single GPU. In the deployment, we set the number of codebooks $C = 16$ and codebook size $S = 256$. Therefore, the number of token embeddings that the Tlow-based sequential model needs to learn is only $C \times S = 4096$, far fewer than the tens of millions of item embeddings in the baseline model. The Tlow-based sequential model is initialized with all parameters of the baseline model except the item-ID embeddings. Before being deployed for online A/B test, both models are trained for two weeks on identical datasets with more than 70 and 200 million interaction records per day in single-domain and cross-domain scenarios, respectively.

\subsubsection{Online Performance}
After deploying the aforementioned Tlow-based and ID-based models as supplemental retrieval pathways, we evaluate the conversion efficacy of both pathways across the entire picture corpus and specifically on cold-start pictures exposed on the same day as publication. The following metrics are reported:
\begin{itemize}[leftmargin=*]
    \item CTR~(Click-Through Rate) measures the click-through rate of items exposed via this specific retrieval pathway.
    \item UCTR~(User Click-Through Rate) averages all the users' click-through rate of items exposed via this pathway.
\end{itemize}

Results are shown in Table \ref{tab:online}. Due to privacy considerations, we report only the relative difference for all metric comparisons between the models. As we can see, the sequential model employing Tlow's tokenization demonstrates $4.79\%$ and $6.23\%$ higher CTR compared to the one using random IDs in single-domain and cross-domain scenarios, respectively. Furthermore, we observe a more substantial improvement of $10.32\%$ and $7.20\%$ in UCTR, indicating its effectiveness in retrieving potentially relevant content to a broader user base. These findings validate Tlow's powerful capabilities on multi-modal and cross-domain embedding transformation in large-scale industrial recommender systems. To further validate Tlow's effectiveness on cold-start recommendations, we perform evaluations exclusively on newly published pictures each day. The results reveal that the advantages of Tlow become markedly more pronounced when evaluated solely on these cold-start pictures, since random ID embeddings require extensive user interaction data for training, whereas Tlow utilizes inherent semantic features available immediately upon a picture's publication.

\begin{table}[!t]
    \centering
    \caption{Online metrics improvement (\%) of pathway comparison in single-domain and cross-domain scenarios.}
    \label{tab:online}
  \setlength{\tabcolsep}{1.2mm}{
    \begin{tabular}{ccccc}
      \toprule
 \multicolumn{1}{c}{\multirow{2.5}{*}{\textbf{Scenario}}} & \multicolumn{2}{c}{\textbf{Overall}} & \multicolumn{2}{c}{\textbf{New Item}}  \\
      \cmidrule(l){2-3} \cmidrule(l){4-5}
       & CTR & UCTR & CTR & UCTR  \\
      \midrule
      Single-domain & 4.79 & 10.32 & 8.46 & 11.64 \\
      \midrule
      Cross-domain & 6.23 & 7.20 & 9.09 & 9.45 \\
\bottomrule
    \end{tabular}}
\end{table}

Beyond pathway comparisons, core online metrics from A/B testing also improve significantly: in the single-domain scenario, 24-hour overall and newly-published picture CTR increase by $0.78\%$ and $1.94\%$, respectively; in the cross-domain scenario, per-capita picture CTR improves by $1.05\%$, accompanied by a $1.15\%$ decrease in top-tier accounts' exposure share, indicating that Tlow substantially benefits long-tail picture exposure. 

\section{Related Work}
\subsection{ID-based Recommendation}
ID-based recommendation originates from collaborative filtering, evolving from matrix factorization~\cite{koren2009matrix} to GNN-based models~\cite{wang2019neural}. For sequential recommendation, various architectures have been explored, including Markov Chains~\cite{rendle2010factorizing}, RNNs and CNNs~\cite{tang2018personalized,hidasi2015session,xu2019recurrent,yan2019cosrec}, and self-attention mechanisms~\cite{zhou2018deep,kang2018self,zhang2025hierarchical}. Multi-modal information such as text and images has also been incorporated to enrich item representations~\cite{zhang2025hierarchical,tao2020mgat}.

\subsection{Tokenization-based Recommendation}
TIGER~\cite{rajput2023recommender} pioneers the use of RQ-VAE to tokenize item semantic embeddings, followed by works that introduce alignment strategies to enhance tokenization quality~\cite{wang2024learnable,liu2024generative,wang2024eager}. However, the hierarchical nature of RQ-VAE leads to low decoding efficiency due to codebook correlations. To enable parallel decoding, PQ-based independent tokenization methods have been proposed~\cite{hou2023learning,hou2025generating}, while parameter-free tokenizers based on heuristic algorithms remain limited in performance~\cite{hua2023index,petrov2023generative,si2024generative}. Despite this progress, existing independent tokenizers still face challenges from non-independent embedding dimensions and complex embedding distributions, which motivates our flow-based approach.

\section{Conclusion}
We propose Tlow, a flow-based item tokenizer that transforms semantic embeddings into a standard normal distribution space, achieving dimensional independence and distributional simplicity for more accurate independent tokenization. Tlow significantly improves recommendation performance across general, cross-domain, multi-modal, and cold-start scenarios in both offline and online experiments.

\begin{acks}
This work is supported by the National Natural Science Foundation
of China under U24B20180.
\end{acks}

\clearpage
\bibliographystyle{ACM-Reference-Format}
\balance
\bibliography{ref}

\end{document}